\documentstyle[12pt]{article}
\newcommand{\beq}{\begin{equation}}
\newcommand{\eeq}{\end{equation}} 

\newcommand{\beqa}{\begin{eqnarray}}
\newcommand{\eeqa}{\end{eqnarray}}

\def\half{\frac{1}{2}}
\def\opone{\leavevmode\hbox{\small1\kern-3.8pt\normalsize1}}

\begin{document}

\title{EPR Test with Photons and Kaons: Analogies}
\author{
N. Gisin\\
\protect\small\em Group of Applied Physics, University of Geneva,  
1211 Geneva 4, Switzerland\\
A. Go \\
\protect\small\em EP Division, CERN, 1211 Geneva 23 Geneva, Switzerland}
\date{\today}

\maketitle
\begin{abstract}
We present a unified formalism describing EPR test using spin 1/2 particles, 
photons and 
kaons. This facilitates the comparison between existing experiments using 
photons and kaons. It underlines the similarities between birefringence and 
polarization dependent losses that affects experiments using optical fibers and
mixing and decay that are intrinsic to the kaons.
We also discuss the limitation these two characteristics impose on the
testing of Bell's inequality.

\end{abstract}

\section{Introduction}
Quantum Mechanics is one of the most successful physics theory
of the 20th century. Over the past 70 years since its inception, Quantum theory
has been tested in many areas of physics with high precision. However,
its radical departure from every day intuition has
troubled physicist for the past 70 years. 
One of the most puzzling prediction is entanglement: the fact 
that a multi-particle wave function automatically implies non-local 
correlation when the constituent particles are spatially separated.
This radical departure from our intuitive understanding of nature has 
prompted hot debates over the years.

\section{The EPR-Bohm Experiment}
In 1935, Einstein, Podolsky and Rosen \cite{EPR} proposed a thought
experiment involving two spatially separated particles by which they 
aimed to illustrate the incompleteness of the then newly developed 
quantum theory (QM). Today their argument is no longer considered to imply
incompleteness of QM, but rather to imply that one of their hidden 
assumption is wrong. 

The original discussion of this thought experiment was in terms of 
momentum and position, but a simpler variant was given
by Bohm \cite{EPRB} using two distant spin $\half$ particles in the singlet
state (Fig. \ref{EPRspin}). 
The spin part of the wave function is given by:

\beq
|\psi\rangle = {1 \over \sqrt{2}} 
        [|+\hat m\rangle_1\otimes  |-\hat m\rangle_2 -
         |-\hat m\rangle_1\otimes  |+\hat m\rangle_2]
\label{singlet}
\eeq
where $|\pm\hat m\rangle_j$ describes a state of the $j^{th}$ particle ($j=1,2$)
 with spin up or down 
along the direction $\hat m$, respectively. 
According to QM, if one measures the spin 
along, say $\hat x$ axis, the outcome is not predetermined by the 
wave function, but whatever result one obtains, the spin of particle 2 will
always be anti-parallel to the spin of particle 1. 

The EPR argument goes as follows:
\begin{itemize}
\item First, the predictions of QM concerning observation of the two 
spatially separated particles are correct.
\item Next, Nature is local. This point was not spelled out precisely,
because it seemed obvious to EPR. It is implied in the sentence ``there is
no longer any interaction.''
The general idea is that one can not influence anything "there"
by acting "here", when "here" and "there" are disconnected, in particular when
"here" and "there" are space-like separated.
\item Thirdly, `If without in any way disturbing a system, 
we can predict with certainty
(i.e., with probability equal to unity) the value of a physical quantity, then
 there exists an element of physical reality corresponding to this physical 
quantity' 
\item Consequently, since one can measure either $S_{\hat x}$ or $S_{\hat y}$ on
particle 1, there are ``elements of reality'' corresponding to 
the physical quantity $S_{\hat x}$ and $S_{\hat y}$ on particle 2.
On the other hand, since QM does not  describe such elements of reality, 
EPR concluded that it must be incomplete.
\end{itemize}

The first point of the EPR argument is a matter of experimental evidence. In
1935 and even when Bell formulated his inequality in 1964 \cite{Bell64} there 
was little evidence that QM correctly describes such two-particle 
distant systems. 
But today there is plenty of evidence in favour of QM.
The third point of the EPR argument received a lot of attention in the vast
literature on the subject. 
However, we view it as a complicated way of stating the
obvious assumption implicit in all natural sciences according to which if one
can predict the result of an experiment with 100\% certainty, then the
system out there has the property corresponding to this predetermined 
result (i.e. if we know the result in advance, Nature knows it also). 
Note that EPR did not assume the existence of "elements of reality" 
(or properties of physical systems, or name it using your
preferred terminology), rather they inferred their existence from the general
concept of locality and by assuming that QM describes distant systems correctly.

At this point, the EPR argument leaves two alternatives open: either QM is
incomplete (as concluded by EPR) or Nature is nonlocal (the only remaining
assumption, point 2 above). However, thanks to Bell's inequality one can
prove  (though there is still a logical loophole, called 
the detection loophole \cite{PhotoEff}) that Nature is nonlocal
(in a precise sense to be described below).
Note that
this does not prove the completeness of QM: it only establishes that the EPR
argument is not valid because its second assumption, the one that Einstein
and his co-authors found so self-evident that they did not spell 
it out explicitly: the locality assumption is wrong. 

The sense in which Nature is nonlocal is not easy to grasp. Let us first
stress that it does not imply any direct conflict with relativity: 
there is no controllable action at a distance, hence no way to use it
for faster than light communication. 
Indeed, according to QM  the result of any measurement on a
quantum system is always independent of any parameter or setting chosen to
measure the other particle. 
Jarrett \cite{Jarrett89} and Shimony \cite{Shimony84} 
have named this important characteristic {\it parameter independence}. 
However the result on one particle may depend on the result obtained on 
the other particle. The difference with classical correlation is that the
result can not be assumed to pre-exist. 
In Jarrett and Shimony terminology this is called {\it outcome dependence}. 
Henceforth we call this kind of non-locality (outcome dependence but parameter
independence) {\bf Quantum non-locality}. 
It is plausible that if Einstein would have realized that quantum non-locality 
does not contradict relativity, he would have revised his
argument, though this will remain questionable for ever.

Many experimental tests have been carried out over the years, 
mostly in two photon experiments with polarization entanglement
\cite{Clauser72,Aspect80s,Weihs98}. 
Recently, one of the author has performed an experimental test with photons 
having time-energy correlation travelling through optical fibers over 
a distance of 10km \cite{Unige}. 
Another test of entanglement has been carried out by the other author 
using massive K-meson  and  the strangeness correlation have been 
measured \cite{CPLEAR}. 
In this paper we will discuss the similarity between kaon anti-kaon correlation
in time evolution and the photon polarization correlation in fibers.

\section{EPR test using Photon Polarization and Optical Fibers}
Most experiments have been done with photons. Hence we recall first the
analogy between the polarization of photons and spin $\half$. 

\subsection{Single photon polarization formalism}
Since any pure
spin $\half$ state is represented by a normalized vector in our familiar
3-dim space,
the set of pure
spin $\half$ states is naturally represented by the unit sphere, called the
Bloch sphere.

Any pure polarization state $|\vec m\rangle$ can also
be represented geometrically by a
point on an abstract sphere called Poincar\'e sphere. The corresponding
normalized vector 
with its associated (normalized) spinor are:
\beq
\vec m=(\sqrt{1-\eta^2}\cos(\varphi),\sqrt{1-\eta^2}\sin(\varphi),\eta)
\eeq
and
\beq
|\vec m\rangle = \left(\matrix{\sqrt{\frac{1+\eta}{2}}e^{-i\varphi/2} \cr
\sqrt{\frac{1-\eta}{2}}e^{+i\varphi/2} } \right)
\eeq
respectively.
The one dimension projector reads:
\beq
P_{\vec m}=|\vec m\rangle\langle\vec m|=\half\left(\matrix{1+\eta &
\sqrt{1-\eta^2}e^{-i\varphi} \cr
\sqrt{1-\eta^2}e^{+i\varphi}  & 1-\eta} \right)
\eeq
Conversely, the vector $\vec m$ is related to the spinor and projector by:
\beq
\vec m = \langle\vec m|\vec\sigma|\vec m\rangle 
       = Tr(\vec\sigma P_{\vec m})
\eeq
where $\vec\sigma$ are the 3 Pauli matrices. The scalar products relate as
follows:
\beq
|\langle\vec m_1|\vec m_2\rangle|^2 = \frac{1+\vec m_1\cdot\vec m_2}{2}
\eeq
Therefore, orthogonal states are represented by opposite points on the
sphere, just as
for spins. Among the continuous infinite number of bases, two are more natural:
the left $|L>$ and right $|R>$ circular polarization states and the vertical
$|V\rangle$ 
and horizontal $|H\rangle$ linear polarization state (where vertical and
horizontal refers
to some characteristic direction in the experimental setup). By convention the
circular states are mapped to the poles of the Poincar\'e sphere, the linear
ones
are then mapped to the equator (we chose conventions avoiding the use of the
imaginary unit i):
\beq
\vec L=(0,0,1)\hspace{1cm} |L>=\frac{1}{\sqrt{2}}\lbrack|V\rangle +
|H\rangle\rbrack
\label{transf1}
\eeq
\beq
\vec R=(0,0,-1)\hspace{1cm} |R>=\frac{1}{\sqrt{2}}\lbrack|V\rangle -
|H\rangle\rbrack
\label{transf2}
\eeq
\beq
\vec V=(1,0,0)\hspace{1cm} |V>=\frac{1}{\sqrt{2}}\lbrack|L\rangle +
|R\rangle\rbrack
\label{transf3}
\eeq
\beq
\vec H=(-1,0,0)\hspace{1cm} |H>=\frac{1}{\sqrt{2}}\lbrack|L\rangle -
|R\rangle\rbrack
\label{transf4}
\eeq

\subsection{Singlet state for photons}
The singlet state (\ref{singlet}) for polarization can now by written
equivalently in
terms of circular or linear polarization states:
\beq
|\psi\rangle = {1 \over \sqrt{2}} [|H\rangle_1 |V\rangle_2  
                                 - |V\rangle_1 |H\rangle_2]
 = {1 \over \sqrt{2}} [|L\rangle_1 |R\rangle_2  
                           - |R\rangle_1  |L\rangle_2]
\label{photonsinglet}
\eeq

For photons travelling in optical fibers, two phenomena 
may affect the polarization correlation: birefringence \cite{bf} 
and  polarization dependent losses (PDL) \cite{PDL}. In practice these
effects can be reduced
down to negligible, but in view of the comparison with the kaon systems, we
shall describe them in the next subsections.

\subsection{Birefringence}
Birefringence is caused by asymmetries in the fibers. These determine a fast
and a slow
polarization modes. Since these modes are usually linear, we identify them
with the 
vertical and horizontal states, respectively.
Birefringence leads to a change of the polarization state.
The Poincar\'e vector $\vec m(z)$ at position z along the optical fiber
rotates around the birefringent axis $\vec\beta$. The angular frequency of
the rotation
depends on the medium and, generally, on the wavelength. For a fixed
wavelength one has:
\beq
\frac{\partial}{\partial z}\vec m(z)=\vec\beta\wedge\vec m(z) \hspace{1cm}
\frac{\partial}{\partial z}|\vec m(z)\rangle =
\frac{-i}{2}\vec\beta\vec\sigma |\vec m(z)\rangle
\label{PMDrot}
\eeq
Accordingly, 
\beq
\vec m(z)=R(\vec\beta(z-z_0))\vec m(z0) \hspace{1cm}
|\vec m(z)\rangle = e^{-i\vec\beta\vec\sigma/2(z-z_0)} |\vec m(z_0)\rangle
\eeq
where $R(\vec\alpha)$ is the rotation matrix around the axis $\vec\alpha$
and angle $|\vec\alpha|$, i.e. the rotation corresponding to the unitary 
operator $U(\alpha)= e^{-i\vec\alpha\vec\sigma/2}$.

Fig. \ref{PhotonSphere} illustrates this rotation in the case of 
linear birefringence for
an axis of rotation $\vec\alpha$ on the equator.
(Also shown on the figure is the trajectory of the Poincar\'e vector 
$\vec m(z)$ when simultaneously affected by polarization dependent loss, an
effect 
that we study in the next subsection.)

The effect of birefringence on the singlet state (\ref{singlet}) is quite
simple. Assuming it
acts only on the second photon, it implies that when the first photon is
measured to be
in a state represented by the Poincar\'e vector $\vec m$, 
then the second photon,
instead of being in the orthogonal state $-\vec m$, is in the "rotated state"
$-R(\vec\beta(z-z_0))\vec m$, thus reducing the polarization correlation. 
However, it is important to notice that this rotated state can be
balanced by appropriately counter-rotating the polarization analyzer. Hence
birefringence in correlation measurements is merely a matter of aligning 
polarization analyzers.

\subsection{Polarization Dependent Loss (PDL)}
Many optical components have polarization dependent losses (PDL), the most
extreme
example being a polarizer that completely attenuates one polarization state
while
not affecting the orthogonal polarization state.
Some special fibers do also have PDL \cite{PDLfiber}. They are made mainly as
polarizing elements for
fiber sensors. The only case we know where such fibers have been used in a
quantum optics
experiment was to demonstrated a potentially useful generalized quantum
measurement \cite{Huttner96}
(i.e. not of Von Neumann type measurement, but an effect
some times called by the horrible name: "Positive Operator Valued
Measurement" or POVM \cite{PeresBook}).

In presence of PDL one state $|+\vec\Gamma\rangle$
undergoes lower losses than its orthogonal state $|-\vec\Gamma\rangle$.
Let $T_{\max}$ and
$T_{\min}$ denote
the maximum and minimum transmission coefficient (for the intensity),
respectively. Then
the evolution operator $T$ reads, in the $|\pm\vec\Gamma\rangle$ basis:
\beq 
T = \pmatrix{\sqrt{T_{\max}} & 0 \cr 0 & \sqrt{T_{\min}}}
\label{PDLstretch}
\eeq
Note that the transmission coefficient for depolarized light is $T_{depol} =
\frac{T_{max} + T_{min}}{2}$.
Arbitrary states $|\vec m>$ evolve as follows:
\beqa
|\vec m\rangle &=& 
\langle+\vec\Gamma|\vec m\rangle\cdot|+\vec\Gamma\rangle
  +\langle-\vec\Gamma|\vec m\rangle\cdot|-\vec\Gamma\rangle \\
&=& P_{+\vec\Gamma}|\vec m\rangle+P_{-\vec\Gamma}|\vec m\rangle \\
&\rightarrow& \sqrt{T_{\max}}P_{+\vec\Gamma}|\vec
m\rangle+\sqrt{T_{\min}}P_{-\vec\Gamma}|\vec m\rangle \\
&=& T|\vec m\rangle
\eeqa
The transmission coefficient are related to the fiber length $z$:
\beq
T_{\max}=e^{-\alpha_{\max}z}\hspace{1cm}T_{\min}=e^{-\alpha_{\min}z}
\eeq
Hence, the above evolution can be see as the solution of an evolution equation:
\beq
\frac{\partial}{\partial z}|\vec
m(z)\rangle=-\left(\half\alpha_{\max}P_{+\vec\Gamma}
+\half\alpha_{\min}P_{-\vec\Gamma} \right)|\vec m(z)\rangle
\eeq
The evolution of the Poincar\'e vector $\vec m(z)$ is more complicated than
merely a
rotation: while remaining on the sphere, $\vec m(z)$ tends towards $|+\vec
\Gamma \rangle$. 
Let the $|\pm \vec \Gamma \rangle$ be ($|L\rangle,|R\rangle$), 
Fig. \ref{PhotonSphere} illustrates such an evolution in the case of 
a birefringent fiber with PDL
(in this example the birefringence axes and the PDL axes $\pm\vec\Gamma$ are
the same, as is the case in real fibers).

The effect of PDL on the singlet state (\ref{photonsinglet}) 
is more subtle than birefringence. Assuming it
acts only on the second photon, it implies that when the first photon is
measured to be
in a state represented by the Poincar\'e vector $\vec m$, then the second
photon,
instead of being in the orthogonal state $-\vec m$, is in a state shifted
toward $+\vec\Gamma$.
Contrary to the case of birefringence, this shifted state can't be
balanced by appropriately setting the polarization analyzer (actually one
possibility would
be to insert in front of the analyzer an element with the same PDL but
opposite axes such that
the two PDL effects result in a polarization independent loss. But without
additional loss
the PDL can't be balanced). Hence PDL
in correlation measurements does destroy irreversibly some of the correlation.

\section{The EPR test using neutral Kaons}
The best example of the massive particle two-state system is the neutral 
Kaon. Neglecting CP violation and
for an initial $J^{PC}=1^{--}$ state, where $J$ is the total angular 
momentun while $P$ and $C$ are parity and charge 
conjugation discrete symmetries,
the wave function can be written in two
different basis (similarly to (\ref{photonsinglet})):
\beq
|\psi\rangle = {1 \over \sqrt{2}} [|K_L\rangle_1  | K_S\rangle_2  -
         | K_S\rangle_1  | K_L\rangle_2],
 = {1 \over \sqrt{2}} [|K^0\rangle_1  | \bar K^0\rangle_2  -
         | \bar K^0\rangle_1  | K^0\rangle_2],
\eeq
where $K_L$ and $K_S$ are eigenstates of weak interaction
while $K^0$ and $\bar K^0$ are eigenstates of strong interaction.
This is in exact analogy to the linear and circular polarization states.
The transformation rules \cite{Perkins86}
between the two above basis are formally identical to the
photon case (\ref{transf1},..,\ref{transf4}), for example:
\begin{equation}
  |K^0\rangle =
  \frac{1}{\sqrt{2}}\lbrack|K_S\rangle +  |K_L\rangle\rbrack  \hspace{1cm}
  |\bar K^0\rangle =
  \frac{1}{\sqrt{2}}\lbrack|K_S\rangle -  |K_L\rangle\rbrack
\label{eq:klk0}
\end{equation}
Geometrically, the $|K_S\rangle$ and $|K_L\rangle$ spinors can be 
represented as points on
the north and south poles on the Poincar\'e sphere,
just as the circular polarization states, while
$|K^0\rangle$ and $|\bar K^0\rangle$ correspond 
to two opposite points on the equator, just as the 
linear polarization state  $|H\rangle$ and $|V\rangle$, see Fig. \ref{KaonSphere}.

\subsection{Mixing and decay of the neutral Kaon}
The kaon time evolution reads:
\begin{equation}
   i\hbar\frac{\partial }{\partial t}|K_S\rangle = 
          (m_S -i\frac {\gamma _S}{2}) |K_S\rangle 
\end{equation}

\begin{equation}
   i\hbar\frac{\partial }{\partial t}|K_L\rangle = 
          (m_L -i\frac {\gamma _L}{2}) |K_L\rangle 
\end{equation}
where $m_L (m_S)$ is the mass of the long (short) eigenstate and
$\gamma_L(\gamma_S) $ is the width of the long (short) eigenstate.
Accordingly, the time evolution operator $U(t)$ is diagonal in $(K_S, K_L)$
and can be
decomposed into a (birefringence-like) rotation and a (PDL-like) contraction:

\beqa
U(t)&=&\pmatrix{ e^{-(im_S+\frac {\gamma _S}{2})t} &  0 \cr
      0 &  e^{-(im_L+\frac {\gamma _L}{2})t} \cr }\\
     &=&e^{-i\frac{m_S+m_L}{2}t}\cdot e^{-i\frac{m_S-m_L}{2}t\sigma_3}\cdot
      \pmatrix{e^{\gamma_St/2} & 0 \cr 0 & e^{\gamma_Lt/2}} \label{decompU}
\eeqa
The first term in (\ref{decompU}) is only a global phase factor. The second
term, with dependence on the mass difference between $K_L$ and $K_S$, produces
mixing between $K^0$ and $\bar K^0$ (also called strangeness mixing), 
an effect formally analog to birefringence. It is represented by
a rotation around the north-south axis of the Poincar\'e sphere, similar to
(\ref{PMDrot}) in Fig. \ref{KaonSphere}.
Finally the third term is due to the fact that neutral kaons are 
unstable particles, $K_S$ decaying faster than $K_L$. 
It is formally identical to the equation  (\ref{PDLstretch})
which describes PDL in fibers. It is represented by the 
precession towards $K_L$ on the Poincar\'e sphere in Fig. \ref{KaonSphere}.

\section{Analogies and differences between the two systems}
The analogies should now be obvious, compare figures \ref{PhotonSphere} and \ref{KaonSphere}.
First, both birefringence in optical fibers
and the strangeness mixing in neutral kaon produce rotations
of the Poincar\'e vectors representing the quantum states of the photons and
kaons, respectively. 
The only (minor) difference is that conventionally one represents 
the rotations axis on the equator for photons while the rotation axis 
for the kaon system passes through the poles
(actually this assumes linear birefringence as in real optical fibers, for
circular birefringence - also called optical activity 
- the rotation axis also passes through the poles.)
Next, PDL in optical fibers like kaon decay produces identical stretches of
the state space. 
As for birefringence, the axes of the stretch differ by convention, but the
resulting effects are identical.

Despite these strong analogies there are also deep differences. 
First, contrary to polarization analyzers, kaon "analyzers" can't be 
simply rotated. Actually, all what can be done is to distinguish 
(i.e. analyze) between the $K^0$ and $\bar K^0$ by their decay or
interaction products. 
In the optical analogy, this means that we have access to only 
one position of linear polarization analyzers.
Experiments with kaons thus need the rotation produced by strangeness mixing
(the apparent birefringence) as a natural way to effectively rotate 
the kaon analyzer by adjusting its distance to the source!
Second, contrary to the PDL of optical fibers, the kaon decay can't be 
reduced nor compensated. 
This is the main drawback of EPR tests using kaons. Indeed, the decay,
contrary to the strangeness mixing, reduces irreversibly the quantum
correlation between the two particles 
(Likewise PDL, but not birefringence, reduces the correlation between the photons.
But, as already mentioned, this correlation can be recovered at the cost
of extra losses). 

Finally, let us also mention the pretty obvious difference: 
kaons, contrary to photons,  are massive particles. 
This important difference by itself justifies the interest in
using kaons for tests of basic quantum mechanics, like entanglement. 
Indeed, it would not be safe to base our confidence in the most peculiar 
prediction of quantum mechanics, like quantum non-locality, 
only on massless particles.

\section{Testing Bell Inequality}
\subsection{The Bell-CHSH inequality}
Very briefly, the Bell-CHSH inequality \cite{CHSH} 
can be derived as follows. Let
$\alpha$, $\alpha'$ and $\beta$, $\beta'$ denote possible settings (i.e.
parameters) of 
Alice's and Bob's measuring devices \cite{AliceBob} (Fig. \ref{EPRspin}). 
Assume the results $\mu=\pm1$ are determined
by the local setting and by a global hidden variable $\lambda$:
$\mu(\alpha,\lambda)=\pm1$ 
and similarly for the other setting. Note the important physical assumption
that the
result on Alice's side does not depend on Bob's setting, and vice-versa:
$\mu(\beta,\lambda)$
is independent of $\alpha$. This is the locality condition in EPR.
The Bell-CHSH inequality is based on the following
trivial inequality (note that either
$\mu(\beta,\lambda)=\mu(\beta',\lambda)$, or
$\mu(\beta,\lambda)=-\mu(\beta',\lambda)$)
\beqa
&&\mu(\alpha,\lambda)\mu(\beta,\lambda)-\mu(\alpha,\lambda)\mu(\beta',\lambda)
+\mu(\alpha',\lambda)\mu(\beta,\lambda)+\mu(\alpha',\lambda)\mu(\beta',\lambda) \nonumber\\
&=&\mu(\alpha,\lambda)\left(\mu(\beta,\lambda)-\mu(\beta',\lambda)\right)
+\mu(\alpha',\lambda)\left(\mu(\beta,\lambda)+\mu(\beta',\lambda)\right) \le2
\label{CHSHlambda}
\eeqa
Define the expectation value of the product of the outcomes on both side for
settings
$\alpha$ and $\beta$:
\beq
E(\alpha,\beta)=\int d\lambda\rho(\lambda)\mu(\alpha,\lambda)\mu(\beta,\lambda)
\eeq
where $\rho(\lambda)$ is a (integrable) probability distribution. The
Bell-CHSH inequality
follows from (\ref{CHSHlambda}):
\beq
S=E(\alpha,\beta)-E(\alpha,\beta')+E(\alpha',\beta)+E(\alpha',\beta')\le2
\label{CHSH}
\eeq

\subsection{Bell tests with photons}
For photon polarization the settings are the angles of linear analyzers with
respect to some arbitrary origin. For the singlet state
(\ref{photonsinglet}), by symmetry one has
$E(\alpha,\beta)=E(|\alpha-\beta|)=-\cos(\alpha-\beta)$ (Fig. \ref{fig:CorFun}a). 
Actually, it suffices to consider the following
one parameters set of settings:
$\alpha=0$, $\alpha'=2\theta$, $\beta=\theta$ and $\beta'=3\theta$.
The Bell-CHSH inequality (\ref{CHSH}) then reads:
\beq
S(\theta)=3E(\theta)-E(3\theta)\le2
\eeq
Fig. \ref{fig:CHSH}a displays the QM prediction of $S(\theta)$, 
There is a clear violation of the Bell-CHSH inequality over the entire 
range $0^o<\theta<68.5^o$ and a maximal violation by a factor $\sqrt{2}$ for
$\theta=45^o$.

In actual experiments, due to the finite efficiency of the detectors, one
does not 
have access to $E(\alpha,\beta)$, only coincidence rates are measurable, like
$R_{++}(\alpha,\beta)$ for the $(+1,+1)$ coincidence rate. Assuming that the
set of detected
events constitute a fair sample \cite{PhotoEff}, one uses the normalized
correlation function
\beq
E_R(\alpha,\beta)=\frac{R_{++}(\alpha,\beta)+R_{--}(\alpha,\beta)-R_{+-}
(\alpha,\beta)-R_{-+}(\alpha,\beta)}
{R_{++}(\alpha,\beta)+R_{--}(\alpha,\beta)+R_{+-}(\alpha,\beta)+R_{-+}
(\alpha,\beta)}
\label{ER}
\eeq 
which is experimentally found to violate the Bell-CHSH inequality 
\cite{Aspect80s,Weihs98,Unige}.

\subsection{Bell tests with kaons}
The situation for kaons is similar except for the already mentioned
intrinsic effective birefringence
(due to strangeness mixing) and the effective PDL (due to kaon decay). Since
the analyzer can
not be rotated, the effective birefringence is used: delaying the
measurement effectively
rotates the analyzer. Hence we denote the settings by times $t_a$ and $t_b$.
This would be perfect if there were no decay. But the latter reduces
coincidence rates \cite{CPLEAR}, e.g.
\beq
R_{++}(t_a,t_b)=\frac{1}{8}e^{-(\gamma_S+\gamma_L)t}\left(e^{-\gamma_s\Delta
t}+e^{-\gamma_L\Delta t}
-2e^{-\gamma\Delta t}\cos[(m_S-m_L)\Delta t]\right).
\eeq
Accordingly the correlations function is (Fig. \ref{fig:CorFun}b):
\beqa
E(t_a,t_b) = - e^{-2\gamma t'} e^{-\gamma \Delta t} \cos[(m_S-m_L)\Delta t]
\label{BellKaonunnorm}
\eeqa
where $\gamma = \frac{\gamma_S + \gamma_L}{2}$, 
$t'={\rm min}(t_a,t_b)$ and $\Delta t={\rm abs}(t_a-t_b)$,
($m_S-m_L\approx 0.477\gamma_S$, $\gamma_S\approx 580\gamma_L$).
This damping makes it impossible to violate the Bell-CHSH inequality 
\cite{Ghiraldi} (Fig. \ref{fig:CHSH}b).
However, if one normalizes the correlation function to the undecayed 
pair of kaons (see (\ref{ER})), then the correlation 
function is less damped (Fig. \ref{fig:CorFun}c) \cite{Fehrs}:
\beqa
E_R(t_a,t_b) = \frac {-2 e^{-\gamma \Delta t} \cos[(m_S-m_L)\Delta t]} 
                   {e^{-\gamma_S \Delta t} + e^{-\gamma_L \Delta t}}.
\label{BellKaon}
\eeqa
It does violate Bell Inequality, though less than the photon case, with a 
maximum of 2.35 instead of $2\sqrt{2}$ (Fig. \ref{fig:CHSH}c).

\section{Conclusion}
The EPR-Bell argument is a beautiful example of a simple reasoning leading to
deep insight into the nature of Physics. It opens the road to experiments
involving distant quantum correlated systems and has triggered fruitful 
philosophical  reflections on our worldview, including proposals to exploit 
entanglement for information processing \cite{QIPIntro98}.

We have shown that experiments with kaons have striking analogies with the most 
studied case of photons. 
The latter have the advantage of being easy to
produce and to propagate over long distances, but they have the drawback that 
they are difficult to detect efficiently. 
Hence, to date all Bell tests using photons are based on the normalized
correlation function $E_R(\alpha,\beta)$ defined in (\ref{ER}). 
Kaon, on the contrary, are massive particles, 
in principle easier to detect by their decay products. 
However, due to their intrinsic decay rate,
one still has to base Bell tests on the normalized correlation function
$E_R(\alpha,\beta)$. 
This decay is formally analog to PDL for photons but it carries 
a very different meaning: while PDL is due to external 
parameters which experimenter has control of (by changing the fiber, for example), 
the decay is an intrinsic property of the kaon which experimenter 
has no control of.
This puts a severe limit for Bell tests, 
it cannot rule out any local theory where the kaon decay is pre-determined.
In other words, such Bell test is not loophole-free. 
The other relevant intrinsic property of kaon, 
strangeness mixing is analog to optical birefringence. Contrary to the decay, 
strangeness mixing is an advantage for Bell tests as it allows to effectively 
rotate the analyzer simply by delaying the measurement.

Finally, let us mention another interesting massive particle system: the
B-meson. A pair of  $B^0 \bar B^0$ created at $\Upsilon(4S)$ resonance 
has exactly the same formalism as the
$K^0 \bar K^0$ but with a major difference: $\gamma_S \approx \gamma_L$. 
As a consequence, the exponential terms in (\ref{BellKaon}) cancel out and 
the correlation function is again sinusoidal with the maximum
again at $2 \sqrt{2}$, just like the photons (Fig. \ref{fig:CHSH}a). 
In fact, such maximal violation of Bell-CHSH inequality could be tested 
experimentally at the asymmetric B-factories: BELLE in KEK, Japan \cite{Belle}
and BaBar in SLAC, USA \cite{BaBar}.

\small
\section*{Acknowledgments}
We thank Maria Fidecaro and Armand Muller for their comments on the paper and
Bruno Huttner for carefull reading of the manuscript. 
This work was partially supported by the Swiss National Science Foundation.

\section*{Figure captions}
\begin{enumerate}
\item Schematics of the entangle pair of spin 1/2 particles with the polarization
analyzers. \label{EPRspin}
\item Poincar\'e Sphere for photon polarization. Double dash line is the
movement of Bloch vector due to birefringence and single dash line is the 
combined effect of Birefringence and PDL 
(state of non absorbed photons; eq. \ref{photonsinglet}). \label{PhotonSphere}

\item Poincar\'e Sphere for neutral Kaons. Double dash line is the
movement of Bloch vector due to strangeness mixing alone and single dash line 
is effect of both strangeness mixing and particle decay (state are 
re-normalized to the undecayed kaons). \label{KaonSphere}

\item Correlation function for 
         a) spin 1/2, polarized photons and Neutral kaon pairs with no decay, 
         b) kaons with decay (eq. \ref{BellKaonunnorm}),
         c) kaons with decay and normalized correlation(eq. \ref{BellKaon}). \label{fig:CorFun}

\item Bell-CHSH Inequality S for a) Spin 1/2 and polarized photons,
         b) Kaons, c) Kaons with re-normalized decay probability.
         The shaded area violates Bell Inequality.
         For B-meson, the curve is almost indistinguisheble to the spin 1/2 
         case a), with $\theta=\frac{\Delta M_B}{\tau_B} 
         \Delta t= 0.723 \Delta t$. \label{fig:CHSH}

\end{enumerate}


\begin{thebibliography}{99}

\bibitem{EPR} {\small A. Einstein, B. Podolsky,N. Rosen, 
              ``Can quantum mechanical description of physical reality 
                be considered complete?''  
               Phys. Rev. {\bf 47}, 777-780 (1935).
           reprinted in  \cite{eprBohr}. }

\bibitem{eprBohr} {\small J. A. Wheeler \& W. H. Zurek, 
                 {\it Quantum Theory and Measurement}, 
                 (Princeton University Press, Princeton, NJ, 1983)}

\bibitem{EPRB}  {\small D. Bohm, {\it Quantum Theory} 
                 (Prentice Hall, Englewood Cliffs, NJ, 1951), pp. 614-622.}

\bibitem{Bell64}  {\small J. S. Bell, 
                  ``On the Einstein Podolsky Rosen paradox''
                   Physics {\bf 1}, 195-200 (1964); 
                   reprinted in \cite{eprBohr}. }

\bibitem{PhotoEff} {\small P. Pearle, 
         ``Hidden-variable Example Based upon data rejection'',
         Phys. Rev. D, {\bf2}, 1418 (1970);

  Ph.H. Eberhard, 
  ``Background level and counter efficiencies required for a loophole-free
  Einstein-Podolsky-Rosen experiment.''
  Phys. Rev. A {\bf47}, R747-R750 (1993).}

\bibitem{Jarrett89} {\small  J. P. Jarrett,  
                {\it Philosophical Consequences of
                Quantum Theory}, eds. J.T. Cushing and E. McMullin,
                (University of Notre Dame Press, Indiana, 1989). }

\bibitem{Shimony84} {\small  A. Shimony, 
                ``Controllable and uncontrollable non-locality'',
                in Proceedings on {\it Foundations of Quantum Mechanics in
                the light of new technology}, ed. Kamefuchi et al.,
                (Physical Society of Japan, 1984) pp. 25-30.}

\bibitem{Clauser72}  {\small J. Freedman, \& J. F.Clauser, ``Experimental
test of local hidden variable theories'', Phys. Rev. Lett. {\bf 28},
938-941 (1972); reprinted in \cite{eprBohr}. }

\bibitem{Aspect80s}  {\small A. Aspect, P. Grangier, \& G. Roger,
``Experimental tests of realistic local theories via Bell's theorem'', 
Phys. Rev. Lett. {\bf 47}, 460-463 (1981); 
A. Aspect, J. Dalibard \& G.
Roger,\ ``Experimental Realization of Einstein-Podolski-Rosen-Bohm
Gedankenexperiment: A New Violation of Bell's Inequalities'', Phys. Rev. Lett. 
{\bf 49 }(2), 91-94 (1982); }

\bibitem{Weihs98}  {\small Weihs, G., Reck, M., Weinfurter, H. \& Zeilinger,
A., ``Violation of Bell's inequality under strict Einstein locality
conditions'', Phys. Rev. Lett. {\bf 81} (23), 5039-5041 (1998).}

\bibitem{Unige} {\small W. Tittel, J. Brendel, H. Zbinden, N. Gisin, 
     Phys. Rev. Lett. {\bf 81} (17), 3563
     (1998); see also Phys. Rev. A {\bf59}, 4150-4163, 1999. }

\bibitem{CPLEAR} {\small A. Go et.al., CPLEAR Collaboration, 
    ``An EPR experiment testing the non-separability of the 
      $\rm{K^0 \bar K^0}$ wave function''
     Phys. Lett. B, {\bf 422}, 339-348 (1998).} 
     

\bibitem{bf} {\small L. Kazovsky, S. Benedetto, A. Willner, 
     {\it Optical Fiber Communication Systems} 
     (Artech House, Boston, MA, 1996).}

\bibitem{PDL} {\small N. Gisin,``Statistics of polarization dependent losses'',
     Optics Communication 114, 399-405 (1995).}

\bibitem{PDLfiber} {\small Snyder A.W. and Ruhl F., 
           ``New s-m single polarization optical fiber'',
                   Electron. Lett. {\bf19}, 185 (1983).}

\bibitem{Huttner96} {\small B. Huttner et al.,  
                     Phys. Rev. A {\bf54}, 3783-3789 (1996).}

\bibitem{PeresBook} {\small A.~Peres, {\it Quantum Theory: Concepts and Methods\/}
       (Kluwer, Dordrecht, 1993) p.~282.}

\bibitem{Perkins86} {\small D. Perkins, 
                     {\it Introduction to High Energy Physics} 
                     (Addison-Wesley, Menlo Park, CA, 1987) 
                     3rd ed., pp. 240-244.}

\bibitem{CHSH} {\small J.F. Clauser, M.A. Horne, A. Shimony, and R.A. Holt, 
       ``Proposed Experiment to test Local hidden-variable theries'',
       Phys. Rev. Lett., {\bf23}, 880-884 (1969);}

\bibitem{Ghiraldi} {\small G.C. Ghiralrdi, R. Grassi, R. Ragazzon,
                {\it Can One Test Quantum Mechanics at the $\Phi$-Factory?} 
                in {\it The DA$\Phi$NE Physics Handbook, } 
                ed. L. Maiani et al, (1992) pp. 283-293.}
\bibitem{Fehrs} {\small M. Fehrs,
                {\it On the Quantum Theory of Measurement},
                Ph.D. thesis, Boston University, (1973), pp. 77-87.}

\bibitem{AliceBob} By convention, Alice and Bob denote the two physicists
               who perform the measurements on the two distant systems. 
               This is commonly used in quantum cryptography.

\bibitem{QIPIntro98}  {\small {\it Introduction to Q computation and
       information}, eds H.K. Lo, S. Popescu \& T.P. Spiller, (World Scientific, 1998)}

\bibitem{Belle} {\small BELLE Collaboration, {\it Technical Design Report},
                {\bf KEK-R-95-1}, (1995).}

\bibitem{BaBar} {\small BaBar Collaboration,
                {\it BaBar Physics Handbook}, {\bf SLAC-R-504},  
                (1998) pp. 717-726.}

\end{thebibliography}
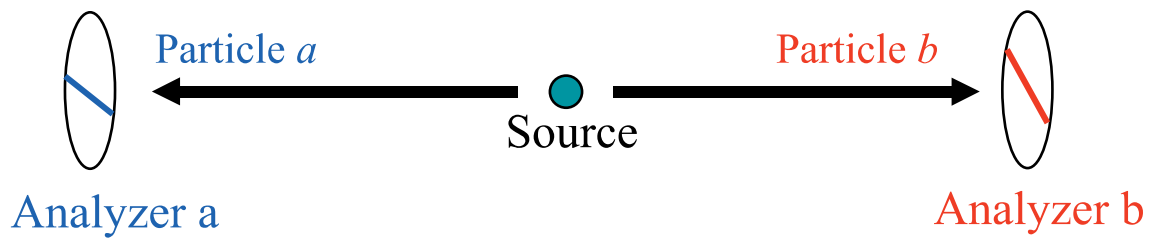
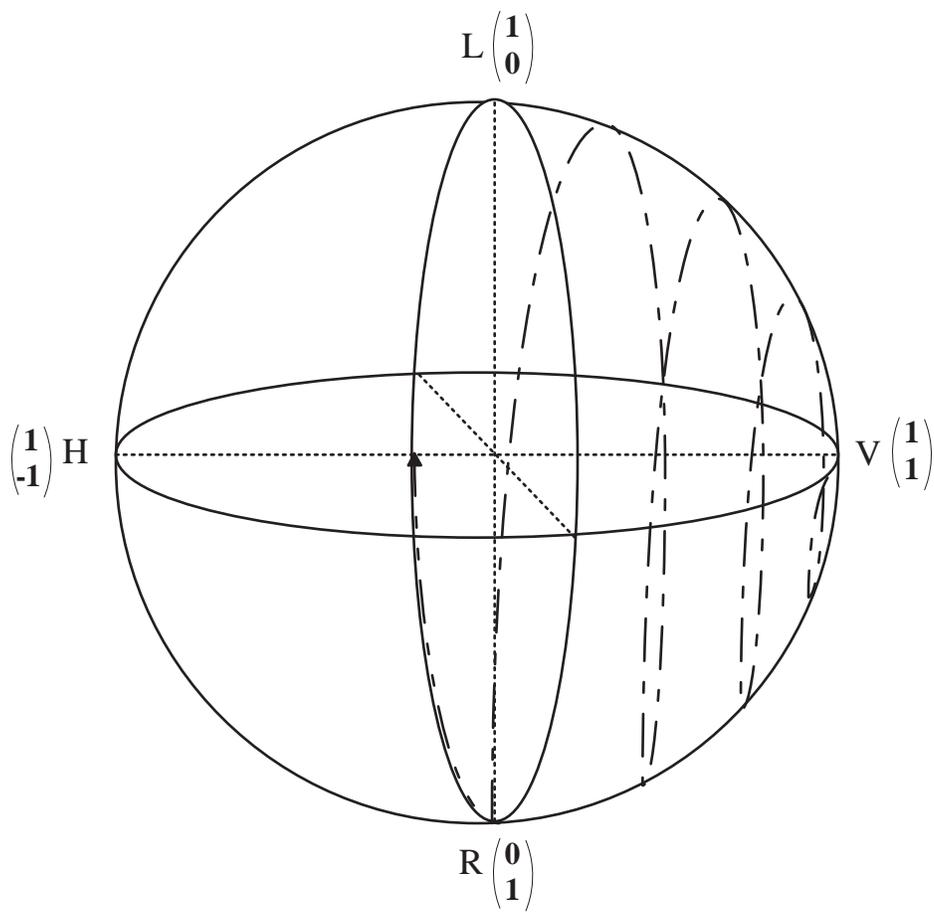
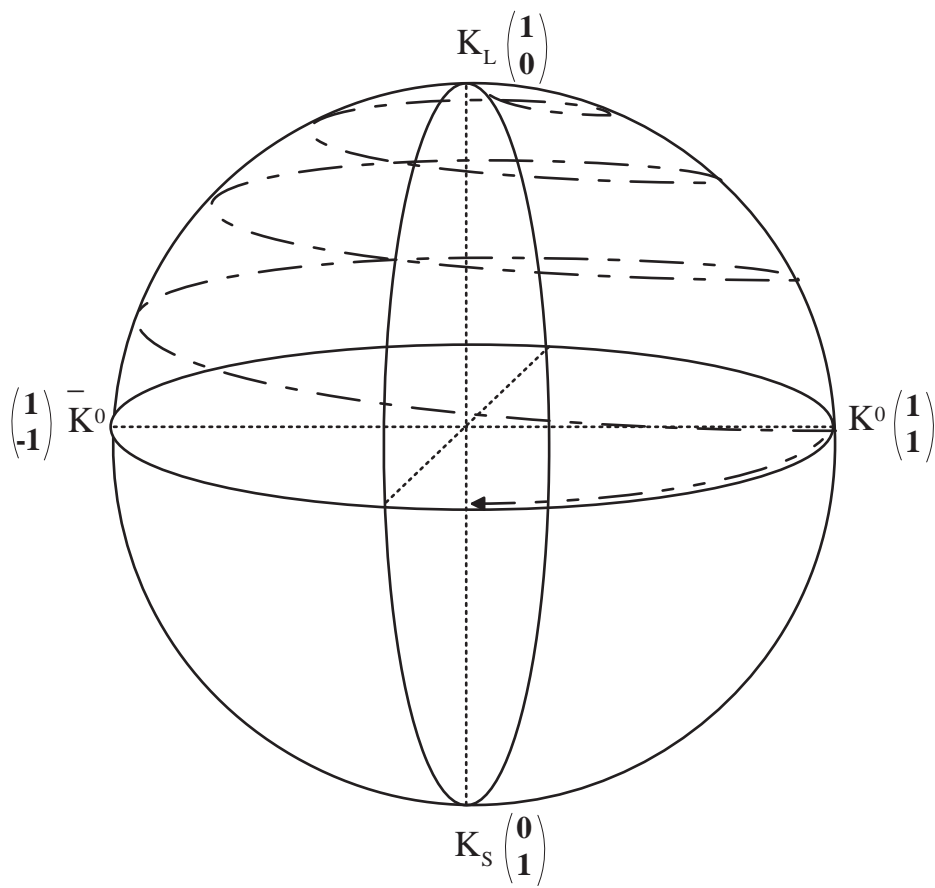
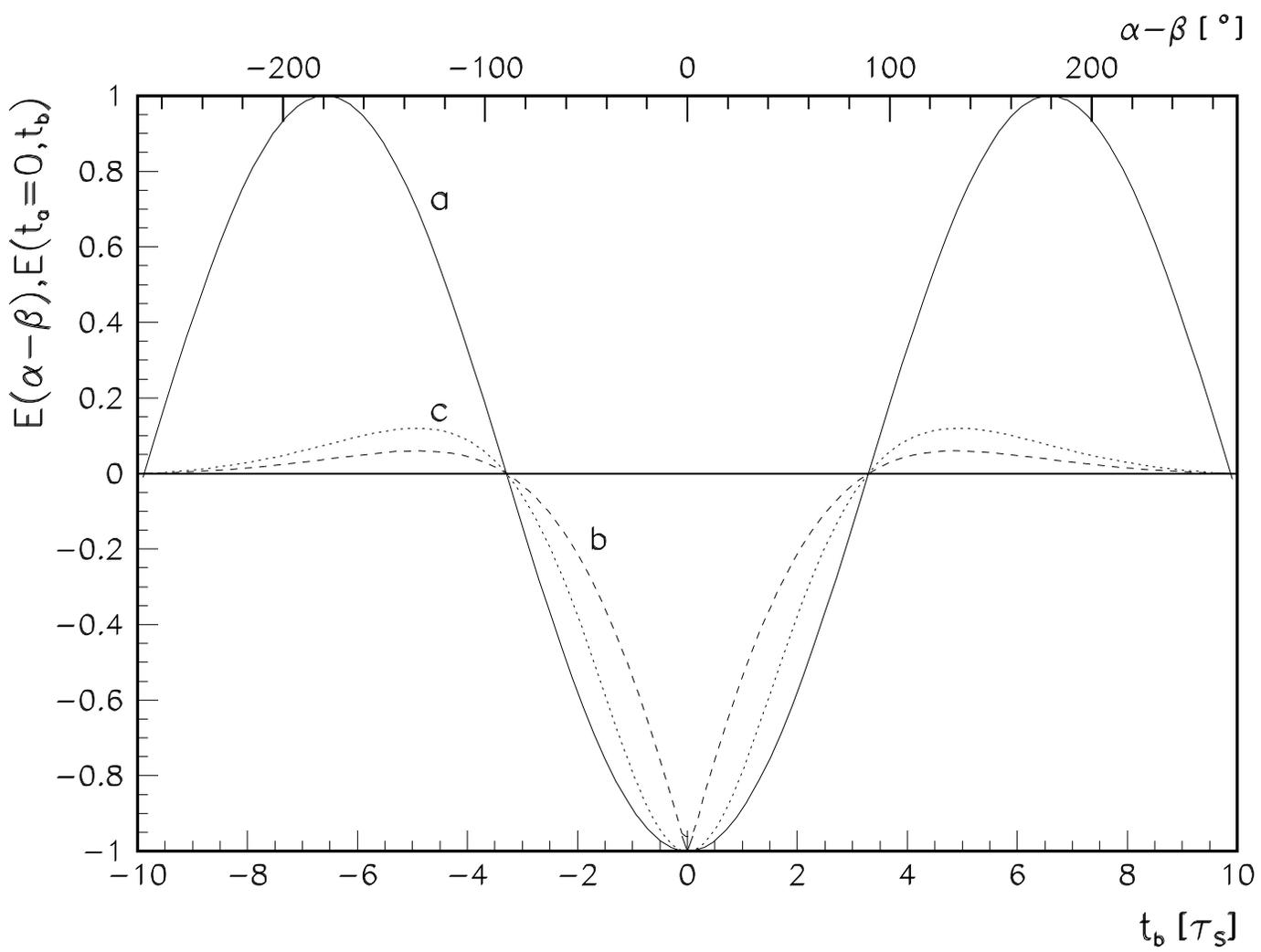
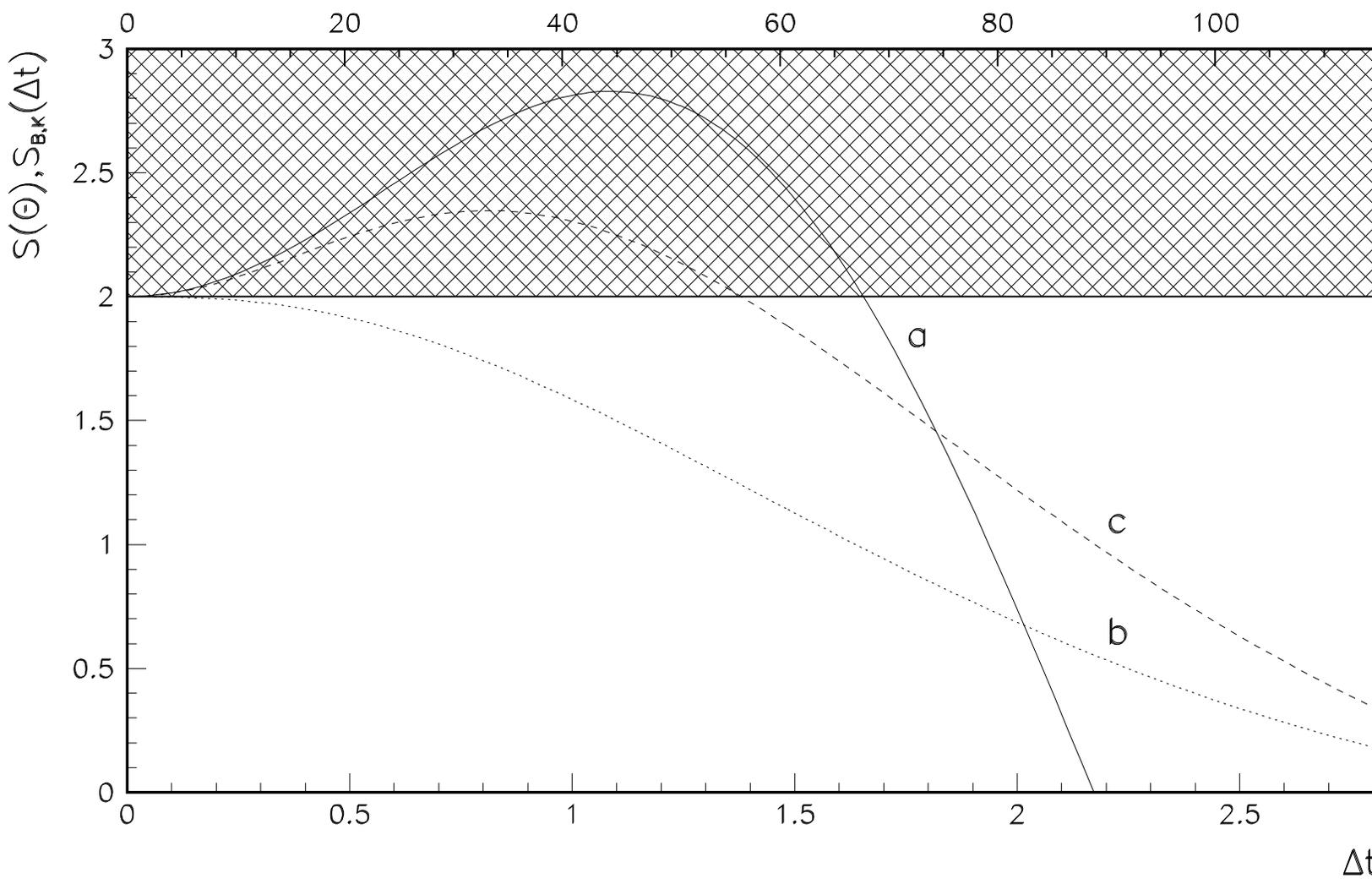
\end{document}